\documentclass[aps,prl,twocolumn,showpacs,superscriptaddress,groupedaddress]{revtex4}  % for review and submission
\usepackage{graphicx}  % needed for figures
\usepackage{dcolumn}   % needed for some tables
\usepackage{bm}        % for math
\usepackage{amssymb}   % for math
\usepackage{color}

\begin{document}

% The following information is for internal review, please remove them for submission
\widetext

\title{Concentration-Dependent Diffusion Instability in Reactive Miscible Fluids}

\author{Dmitry Bratsun} \affiliation{Theoretical Physics Department, Perm State Humanitarian Pedagogical University, 614990, Perm, Russia}
\author{Konstantin Kostarev}
\author{Alexey Mizev}
\author{Elena Mosheva}
\affiliation{Institute of Continuous Media Mechanics, Acad. Koroleva st. 1, 614013, Perm, Russia}

%\date{\today}

\begin{abstract}
We report new chemoconvective pattern formation phenomena observed in a two-layer system of miscible fluids filling a vertical Hele-Shaw cell. We show both experimentally and theoretically that the concentration-dependent diffusion coupled with the frontal acid-base neutralization can give rise to formation of the local unstable zone low in density resulting in a perfectly regular cell-type convective pattern. The described effect gives an example of yet another powerful mechanism which allows the reaction-diffusion processes to govern the flow of reacting fluids under gravity condition.
\end{abstract}


\maketitle

In last decades the interaction between reaction-diffusion phenomena and pure
hydrodynamic instabilities has attracted increasing interests both
from the fundamental point of view of nonlinear science and from
the chemical engineering \cite{avnir95,pons00,eckert08}.
The interest arises from the fact that the chemically-induced changes
of fluid properties such as density, viscosity, thermal conductivity or
surface tension may result in the instabilities,
which exhibit a large variety of convective patterns.

Scenario for instability development essentially differs for immiscible and 
miscible systems of liquids.
The simple, irreversible chemical scheme such as a neutralization
reaction A+B$\to$S occurring in binary liquid-liquid immiscible systems
was studied in~\cite{eckert99,eckert04,brat04,brat08,brat11,brat14}.
The pattern formation in the form of irregular plumes 
and fingers was shown to originate from the coupling
between different gravity-dependent hydrodynamic instabilities
occurring when an organic solvent containing an acid A 
is in contact with an aqueous solution
of an inorganic base B~\cite{eckert99}.
This irregularity looked natural since the configuration of more dense acid on top
of less dense base in the presence of gravity is unstable via the Rayleigh-Taylor (RT)
mechanism~\cite{brat08,brat11}.~A much more impressive, regular pattern of cellular-like fingers 
keeping contact with the interface
was reported for an organic base~\cite{eckert04}.
The regularity was shown to originate from the perfect balance
between the RT instability on the one hand, and 
the Rayleigh-B\'enard~\cite{brat14} and Marangoni~\cite{brat04} mechanisms on the other. 
Thus, a liquid-liquid interface was recognized to be important for performing 
fine-tuning of salt fingers.

A completely different situation was observed in the miscible case.
The main engine breaking the equilibrium here was found to be the difference between 
the diffusion rates of all three substances resulting in double diffusive (DD) instability or 
diffusive-layer convection (DLC) as well as RT instability~\cite{trevel}. 
All the works devoted to this subject at a given moment, usually noted the formation
of irregular patterns of fingers. For example, recently
Almarcha {\it et al.}~\cite{alma10,alma11,carb13}
have shown that the various possible convective regimes can be triggered by
acid-base reactions when a less dense acid solution lies on top of a denser
alkaline one in the gravity field. The possible dynamics are a composition
of only two asymptotic cases: irregular plumes induced by a local RT instability 
above the reaction zone and irregular fingering in the lower solution
induced by differential diffusive effects.

Absolutely in all works in this field, cited or not, the diffusion coefficients 
of species have been assumed to be constant.~Generally, a concentration-dependence 
exists in most systems, but often, 
e.g. in dilute solutions, the dependence is weak and the diffusion coefficient 
can be assumed constant~\cite{crank}. 
This is especially true for fluid mechanics~\cite{bat}. Some rare examples 
of the influence of concentration-dependent diffusion include  
the colloid ultrafiltration~\cite{ri} and membrane transport~\cite{ash} where 
the basic fluid flow is just slightly modified.
Reaction-diffusion problems include the plasma wave dynamics~\cite{ro}
and the Turing instability under centrifugal forces~\cite{guiu}. 

In this Letter, we also focus on the study of chemo-hydrodynamic processes 
which accompany a frontal neutralization reaction taking place between 
two miscible liquids.~We report a new type of instability, 
{\it the concentration-dependent diffusion} (hereinafter CDD) instability
from the family of the double-diffusion phenomena~\cite{tur}. 
It arises when the diffusion coefficients of species depend 
on their concentrations.~We demonstrate both experimentally and theoretically that 
chemically-induced changes of reagent concentrations coupled with 
concentration-dependent diffusion can produce spatially localized 
zone with unstable density stratification that under gravity gives rise 
to the development of perfectly periodic convective structure even in a miscible system. 

%%%%%%%%%%%%%%%%
%  FIGURE 1
%%%%%%%%%%%%%%%%
\begin{figure*}
\includegraphics [scale=0.72] {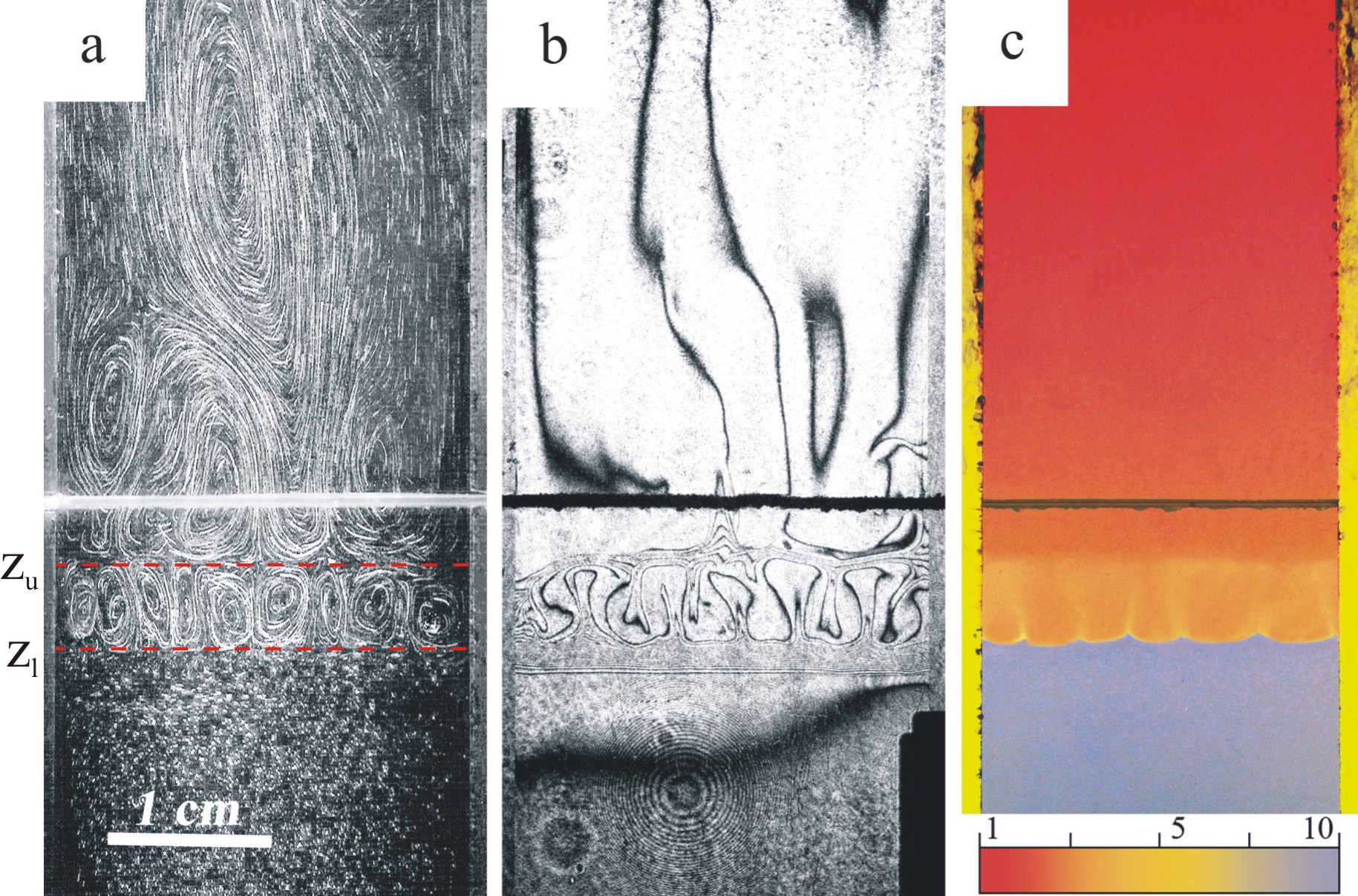}\quad
\includegraphics[viewport=0 0 215 260,clip,scale=0.85]{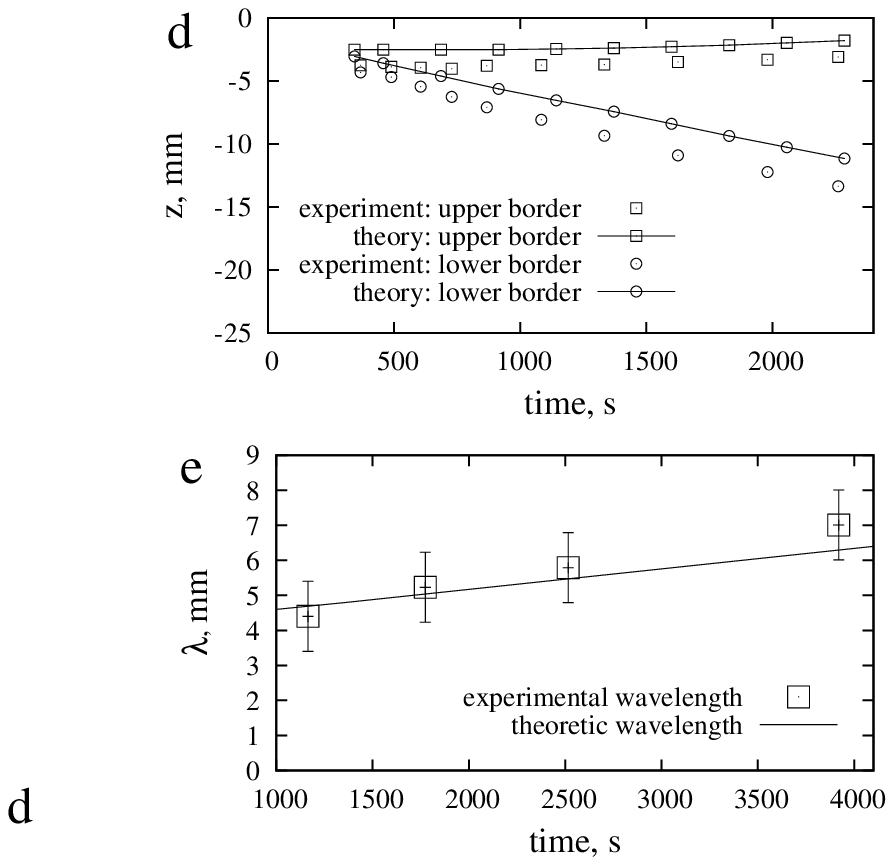}
\caption{\label{fig:1} 
(a-c) Chemo-convective structures arising due to the CDD instability 
observed $1100$ s after as the aqueous solutions 
of HNO$_3$ (top) and NaOH (bottom) were brought into contact in a vertical Hele-Shaw cell: 
(a) Velocity field revealed by the tracks of light-scattering particles; 
(b) Interferogram showing a refractive index distribution; 
(c) {\it p}H distribution obtained in the presence of color indicator. 
Initial concentrations of species are both equal to 1 mol/l. 
The initial contact line is indicated by the horizontal band. 
(d,e) Evolution of the upper (indicated in (a) as Z$_u$) and lower (Z$_l$) 
boundaries (d) and wavelength (e) of the CDD pattern in time obtained experimentally (points) 
and numerically within the theoretical model (lines).}
\end{figure*}

%
%	experiment

{\it Experimental results}. The experiments were performed in a vertically oriented 
Hele-Shaw cell made of two glass plates (width 2.5 cm $\times$ height 9.0 cm)
separated by a thin gap of 0.12 cm. The cell was filled with aqueous solutions of reagents 
whose concentration always provided a steady stratified density distribution. 
We examined a few acid-base pairs formed by HCl or HNO$_3$ 
from one side and NaOH or KOH from another.~During the filling of the cell 
with the upper solution, the lower layer was separated 
by a thin plastic slide inserted in two narrow (0.3 mm) slots made in the walls. 
Fizeau interferometry was used to visualize a refractive index distribution. 
The latter was caused by inhomogeneities induced by the concentration distribution of species
and reaction exothermicity.
The maximum temperature increase measured in the vicinity of the reaction front 
was found to be near 1~K. 
In this case the refractive index deviation due to temperature was at least one order 
of magnitude smaller than that caused by concentration. Thus, the interferograms 
obtained in the experiments reflected mainly the concentration distribution. 
Silver-coated hollow glass spheres were added to the liquids to observe 
the convective patterns which form during the reaction. 
We visualized also the {\it p}H distribution by adding a small amount of universal 
acid-base indicator to the solutions. The comparison of the results obtained 
with and without the indicator has shown that the presence 
of the indicator did not influence the instability scenario and pattern 
formation process, as it was demonstrated in some recent studies~\cite{alma10b,kuster}. 

Right after the prepared solutions were brought into contact, 
the transition zone started to form between them where  
the reagents were transported towards the reaction front only via the diffusion mechanism. 
Then the occurrence of a depleted layer just above 
the diffusion zone has given rise to the formation of plumes which result 
in the development of weak buoyancy-driven convection in the whole upper 
layer. A few minutes after the beginning of the experiment, 
the fluid flow in the form of the periodic array of convective cells (Fig.~\ref{fig:1}a) 
has been formed within the diffusion zone just above the reaction front. 
The cells were arranged between two areas of immobile fluid 
which definitely indicated the formation of a local ``pocket'' with the unstable density stratification
(Fig.~\ref{fig:1}b). 
One can note that the cellular structure did not interact 
with the convection in the upper layer. 
The observations of {\it p}H distribution (Fig.~\ref{fig:1}c) have shown that above 
and below the cells band the medium has almost homogeneous {\it p}H (acidic or alkaline). 
The {\it p}H within the cellular pattern is more neutral indicating the accumulation 
of the reaction product in this zone. Clearly defined zone of the intermediate acidity 
suggests that we deal here with some kind of a cooperative phenomenon.
\begin{table}[b]
\caption{\label{table1} Initial concentrations ratio $\mu$ of different acid-base pairs 
for which the cellular chemostructure has been observed.}
\begin{ruledtabular}
\begin{tabular}{cccc}
\textrm{Acid:Base}&
\textrm{HNO$_3$:NaOH}&
\textrm{HNO$_3$:KOH}&
\textrm{HCl:NaOH}\\
\colrule
$\mu$ & 1 : 1 & 1 : 1.3 & 1 : 0.7
\end{tabular}
\end{ruledtabular}
\end{table}

We found that the structure can exist 
for several hours with the band slowly widening with time (Fig.~\ref{fig:1}d) 
that results in the wavelength growth (Fig.~\ref{fig:1}e). 
It is important to note that this chemo-convective regime was found also
in all pairs of reactants used in experiments, but only at certain ratio 
of initial concentrations (Table 1).

%
%	theory

{\it Theoretical model}.~To describe the observed phenomenon of the CDD instability 
consider two miscible fluids filling a close parallelepiped sufficiently
squeezed along one horizontal direction to use a Hele-Shaw 
approximation~\cite{brat11}.
The upper and lower layer are aqueous solutions of acid $A$ and base $B$ respectively.
Right after the process starts, the acid and base diffuse into each other and are
neutralized with the formation of salt $S$ with the rate $K$. 
The system geometry is given by two-dimensional domain
with $x$-axis directed horizontally and $z$-axis
anti-directed to gravity. We choose the following characteristic scales: 
length - the gap-width $h$, time - $h^2/D_{a0}$, velocity - $D_{a0}/h$, 
pressure - $\rho_0\nu D_{a0}/h^2$ and concentration - $A_0$.
Here $D_{a0}$, $\rho_0$, $\nu$, $A_0$ define constant acid diffusivity, solvent density,
kinematic viscosity and initial acid concentration respectively.
The mathematical model we develop consists in the set of reaction-diffusion-convection
equations coupled to Navier-Stokes equation, 
written in the dimensionless form:
\begin{eqnarray}
\Phi=-\nabla^2\Psi ,
\label{eq:1}
\\
\frac{1}{Sc}\left(\partial_t \Phi+\frac{6}{5}J(\Psi,\Phi)\right) =
\nabla^2\Phi - 12\Phi  - R_a \partial_x A -\nonumber
\\
 - R_b \partial_x B - R_s \partial_x S ,
\label{eq:2}
\\
\partial_t A + J(\Psi,A) = \nabla D_a(A) \nabla A - \alpha AB ,
\label{eq:3}
\\
\partial_t B + J(\Psi,B) = \nabla D_b(B) \nabla B - \alpha AB ,
\label{eq:4}
\\
\partial_t S + J(\Psi,S) = \nabla D_s(S) \nabla S + \alpha AB ,
\label{eq:5}
\end{eqnarray}
where $J$ stands for the Jacobian determinant $J(F,P)\equiv 
\partial_z F\partial_x P - \partial_x F\partial_z P$.~Here 
we use a two-field formulation for motion equation, and introduce the stream
function $\Psi$ and vorticity $\Phi$ defined by~(\ref{eq:1}).
Eq.~(\ref{eq:2}) differs from a standard Navier-Stokes equation
by the additional term $12\Phi$ which is responsible for the average friction force
due to the presence of the side-walls. Diffusion terms in Eq.~(\ref{eq:3}-\ref{eq:5})
have been written in the most general form~\cite{crank}.

The problem parameters are the Schmidt number $Sc=\nu/D_{a0}$,
the Damk\"ohler number $\alpha = K A_0 h^2/D_{a0}$
and the set of solutal Rayleigh numbers for species
$R_i = g\beta_i A_0 h^3/\nu D_{a0}$, $i=\{a,b,s\}$.~Their 
values for the pair HNO$_3$ / NaOH (see Table 1) have been estimated as follows:
$Sc=10^3$, $\alpha=10^3$, $R_a = 1.5\times 10^3$, $R_b = 1.8\times 10^3$, 
$R_s = 2.4\times 10^3$ (see Supplementary information B).

The boundary conditions for Eqs.~(\ref{eq:1}-\ref{eq:5}) are
\begin{eqnarray}
\Psi=0,\quad\partial_i\Psi =0,\quad\partial_i A =0,\quad\partial_i B =0,\quad\partial_i S =0 ,
\label{eq:7}
\end{eqnarray}
where $i=\{x,z\}$ for side-walls and horizontal boundaries respectively. The initial conditions at $t=0$ are
\begin{eqnarray}
z\le 0:\qquad \Psi=0,\quad \partial_z\Psi =0,\quad A=0,\quad B=1; \nonumber
\\
z>0:\qquad \Psi=0,\quad \partial_z\Psi =0,\quad A=1,\quad B=0.
\label{eq:8}
\end{eqnarray}
%

%%%%%%%%%%%%%%%%
%  FIGURE 2
%%%%%%%%%%%%%%%%
\begin{figure}
\includegraphics[viewport=0 0 210 200,clip,scale=0.56]{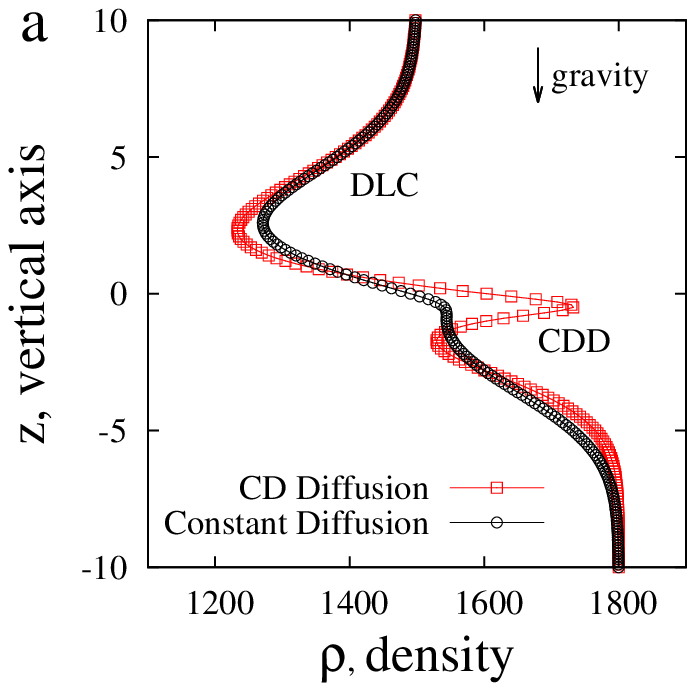}\quad
\includegraphics[viewport=0 0 210 200,clip,scale=0.56]{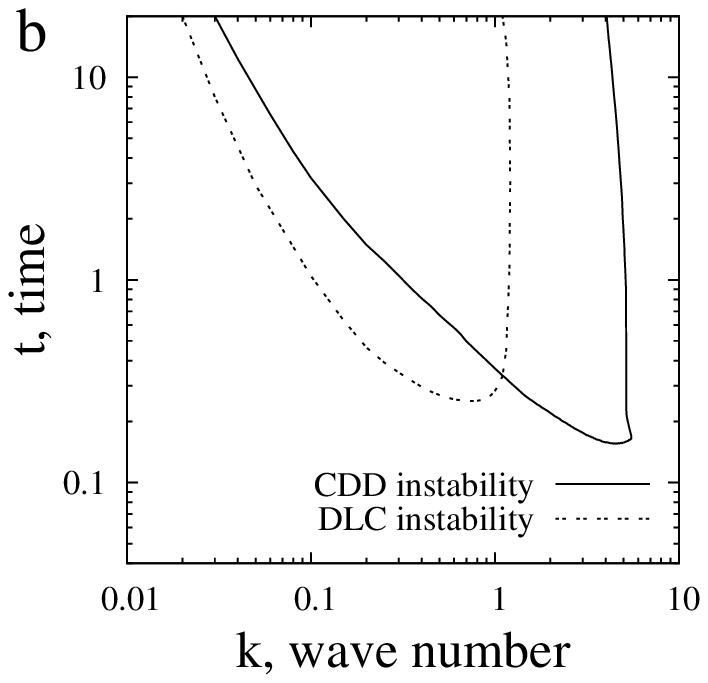}
\caption{\label{fig:2}  (a) Instantaneous base state profiles of the 
total density~(\ref{eq:10}) for the case of constant diffusion (circles) and
concentration-dependent diffusion (squares) at $t=5$;
(b) Neutral curves for DLC and CDD instabilities which arise in two zones low in density 
shown in Fig.~\ref{fig:2}a.}
\end{figure}

We found that a concentration-dependence of diffusion 
plays an important role in the pattern formation. 
Thus, in Eq.~(\ref{eq:3}-\ref{eq:5}) the diffusion coefficients  
have been assumed to be not constant, but depend on their own concentration:
$D_a(A)$, $D_b(B)$ and $D_s(S)$. In order to evaluate the diffusion formulas
for the pair HNO$_3$/NaOH, we have brought together all the known to us experimental data
(see Supplementary information A) and have obtained
\begin{eqnarray}
D_a(A) \approx 0.158 A + 0.881 ,\nonumber
\\
D_b(B) \approx - 0.087 B + 0.594 ,
\label{eq:9}
\\
D_s(S) \approx - 0.284 S + 0.478 .\nonumber
\end{eqnarray}

It follows from~(\ref{eq:9}) that 
the salt is most immobile, compared with acid and base.
In addition, the diffusivity of salt decreases with the growth of the salt concentration.
All these factors together produce an interesting effect.
In order to describe it in term of the buoyancy it is convenient to introduce
the total dimensionless density:
\begin{equation}
\rho (x,z) = R_a A(x,z) + R_b B(x,z) + R_s S(x,z).
\label{eq:10}
\end{equation}

The base state profiles for the density~(\ref{eq:10}) are shown in Fig.~\ref{fig:2}a
for two different diffusion laws: (i) the constant diffusion with standard table values for coefficients;
(ii) the concentration-dependent diffusion defined by~(\ref{eq:9}). 
We see that the curve has only one minimum above the reaction front 
in the case (i) versus two minima (above and below the reaction front) in the case (ii).
The lower minimum enclosed within the regions with a stable stratification
occurs because of the progressively slower diffusion of the salt resulting
in its accumulation in or near the reaction front.
Both minima enable for the potential development of the instability
in the presence of gravity. The upper minimum is the typical for the DLC instability 
(Fig.~\ref{fig:2}a, squares).~The lower local minimum on the same curve is much more interesting:
since it has appeared exclusively due to the concentration-dependence of diffusion, 
we have named it as the CDD instability.

A nonsteady spectral amplitude problem has been solved by the method 
suggested originally in~\cite{homsy86} and developed for chemo-convection in~\cite{brat04,brat14}.~Fig.~\ref{fig:2}b shows 
the neutral curves for the DLC and CDD instabilities. 
At time $t\approx 0.15$ the CDD disturbance with a wave number $k\approx 4.6$
is the first to lose stability. Then more and more waves are involved into the instability area.
The DLC instability arises at  $t\approx 0.25$, and its critical wave number
at the very beginning is $k\approx 0.75$. 
Eventually the maximum growth rate of disturbances in the instability balloon 
is shifted towards longer wavelengths.

%%%%%%%%%%%%%%%%
%  FIGURE 3
%%%%%%%%%%%%%%%%
\begin{figure}
\includegraphics[viewport=40 27 310 130,clip,scale=0.85]{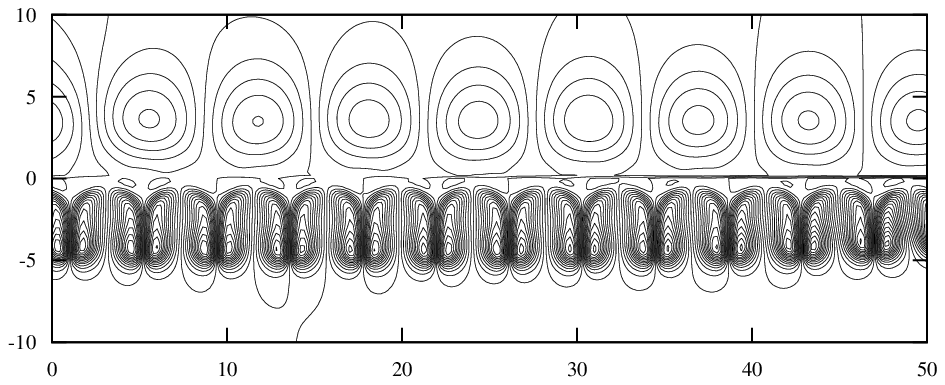}
\includegraphics[viewport=40 22 310 128,clip,scale=0.85]{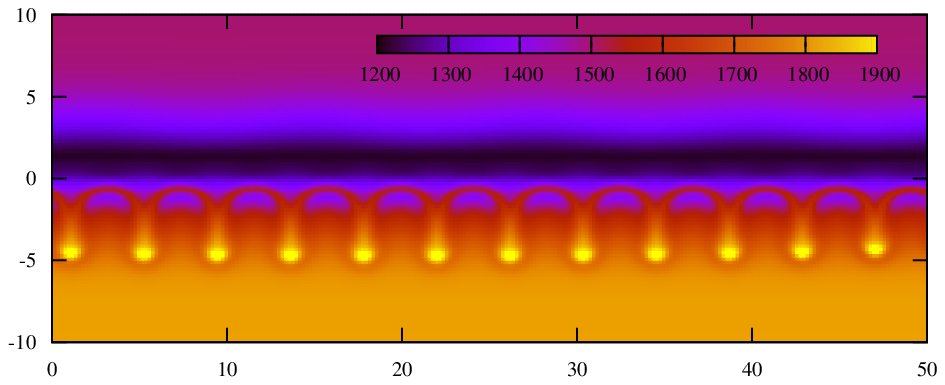}
\caption{\label{fig:3} Stream function (top) and total density (bottom)
obtained by numerically solving full non-linear set of equations (\ref{eq:1}-\ref{eq:9}) for time $t=3$.
The line $z=0$ corresponds to the initial contact line between layers. Non-linear development 
of the CDD instabilitiy below the contact line is clearly seen.}
\end{figure}

To see non-linear development of the disturbances,
the problem (\ref{eq:1}-\ref{eq:9}) has been solved numerically by a finite-difference
method described~\cite{brat14} in detail.
Stream lines and total density of the pattern at time $t=3$ are presented
in Fig.~\ref{fig:3}.~We found that in both zones low in density
the convection develops independently (Fig.~\ref{fig:3}, top).
The most interesting situation is in the lower area where
the cellular chemoconvection with a perfectly periodic structure induced by the CDD instability
has been observed. 
As in the experiment, the boundaries of the structure slowly move apart with time (Fig.~\ref{fig:1}e). 
At $t=3$ the pattern wavelength is about $4.2$, which 
is in good agreement with the experimental data (Fig.~\ref{fig:1}d).

{\it Discussion and closing remarks}. -- The system of miscible fluids when a given 
solution is placed above a denser solution
with the fastest diffusing species in the upper layer has been classified in~\cite{trevel} 
as a typical case of the DLC instability. It occurs due
to the formation of a depleted zone low in density which develops above
the initial contact line while an accumulation zone
where the density is maximal is obtained below the line. As a result, 
irregular DLC fingering is observed which develops on both sides of the initial 
contact line. In our case we also can identify the DLC plumes above 
initial reaction front (see the upper zone in Fig.~\ref{fig:1}a and Fig.~\ref{fig:3}).
But below the contact line we meet a new kind of instability. 
It may occur only in the reactive case when an emerging component 
starts to accumulate near the reaction front. 
If its molecules quickly leave the reaction zone, then it has no significant influence 
on the instability scenario. 
But if the diffusion coefficient of the reaction product decreases with growth 
of its concentration (the CDD effect), it can progressively make a local minimum 
in the density profile (figuratively, ``density pocket''). 
Finally, under gravity condition one can observe the development of the 
localized cellular convection within the bulk of the motionless liquid 
(see the lower zone Fig.~\ref{fig:1}a and Fig.~\ref{fig:3}).

After the localized CDD structure occasionally 
was found in the pair HNO$_3$/NaOH, we have tested a number of other systems and 
found a similar patterning there. In our opinion, it may indicate that the discovered 
effect is of a general nature and should be taken into account 
in reaction-diffusion-convection problems as another tool to organize 
the movement of the reacting fluids.~The CDD effect should take its place 
among other instabilities (DD, DDD, DLC) of the family of the double-diffusive phenomena 
introduced in physics over 60 years ago~\cite{tur}.

We wish to thank A. De Wit for stimulating discussions. 
The work was supported by the Perm Education Ministry, 
RFBR (13-01-00508a, 14-01-96021r\_ural\_a) and Basic research program of UBRAS (\#15-10-1-16).

\end{document}